*Approved for public release; distribution is unlimited.*

# DCL System Using Deep Learning Approaches for Land-Based or Ship-Based Real-Time Recognition and Localization of Marine Mammals

**Peter J. Dugan**
Bioacoustics Research Program
Cornell Laboratory of Ornithology
Cornell University
159 Sapsucker Woods Road, Ithaca, NY 14850

phone: 607.254.1149 fax: 607.254.2460 email: pjd78@cornell.edu

**Christopher W. Clark**
Bioacoustics Research Program
Cornell Laboratory of Ornithology
Cornell University
159 Sapsucker Woods Road, Ithaca, NY 14850

phone: 607.254.2408 fax: 607.254.2460 email: cwc2@cornell.edu

**Yann André LeCun**
Computer Science and Neural Science
The Courant Institute of Mathematical Sciences
New York University
715 Broadway, New York, NY 10003, USA

phone: 212.998.3283 mobile phone: 732.503.9266 email: yann@cs.nyu.edu

**Sofie M. Van Parijs**
Northeast Fisheries Science Center, NOAA Fisheries
166 Water Street, Woods Hole, MA 02543

phone: 508.495.2119 fax: 508.495.2258 email: sofie.vanparijs@noaa.gov

http://www.birds.cornell.edu/brp

## LONG-TERM GOALS

The ONR DCL grant focuses on advancing state-of-the-art of data-mining for the bioacoustics community through researching and creating new technologies, algorithms and systems to decipher and understand very large passive acoustic datasets. The long-term goal is to develop advanced computational systems and algorithms that will provide scientists the ability to efficiently harvest animal vocalizations from large, complex datasets. The newly developed systems provides efficient, high performance processing of acoustic sounds by allowing a state-of-the-art technology to host algorithms for advanced detection and classification and other data-mining strategies.

## OBJECTIVES

While the animal bioacoustics community at large is collecting huge amounts of acoustic data at an unprecedented pace, processing these data is problematic. Currently in bioacoustics, there is no effective way to achieve high performance computing using commericial off the shelf (COTS) or government off the shelf (GOTS) tools. Although several advances have been made in the open source and commercial software community, these offerings either support specific applications that do not integrate well with data formats in bioacoustics or they are too general. Furthermore, complex algorithms that use deep learning strategies require special considerations, such as very large libraiers of exemplars (whale sounds) readily available for algorithm training and testing. Detection-classification for passive acoustics is a data-mining strategy and our goals are aligned with best practices that appeal to the general data mining and machine learning communities where the problem of processing large data is common. Therefore, the objective of this work is to advance the state-of-the art for data-mining large passive acoustic datasets as they pertain to bioacoustics.

## APPROACH

The following research occurred between 2012 and 2015, with a preliminary three month period in 2011. The 2011 three month preliminary design phase involved the team collaborating and settling on a framework for the plan to advance state-of-the-art data-mining of passive acoustic sound archives. Results of the planning divided the work into three stages: system development, data-mining algorithms research, and collaborative projects. Although the grant was intended to explore aspects of deep-learning as they relate to bioacoustics, the initial planning suggested from the beginning that bioacoutsics did not yet have adequate systems to support large amounts of data required for deep recognition and other advanced technologies. With this basic deficiency recognized at the forefront, portions of the grant were dedicated to fostering deep-learning by way of international competitions (kaggle.com) meant to attract deep-learning solutions. The focus of this early work was targeted to make significant progress in addressing big data systems and advanced algorithms over the duration of the grant from 2012 to 2015. This early work provided simulataneous advances in systems-algorithms research while supporting various collaborations and projects. Advances in the grant were also planned to develop from needs of the actual contracts and analysis projects.

## WORK COMPLETED

### PRELIMINARY WORK

Discussions summarizing the state of advanced machine learning revealed several key foundational aspects. First, groups like NYU were using (1) large datasets for training and testing; (2) had systems to deal with running large amounts of data for training, testing and validating algorithms; (3) collaborated using open source tools, competitions and conferences for sharing and standardizing; and (4) suggested that BIG Data was the next generation for many fields. It was obvious that NYU (and other labs) had been working toward this Big Data horizon by developing specialized software, such as LeNet or (Lush), which may not be easily integrated into the marine mammal research community. The forward plan for Phase II suggested four key areas on which to focus: (1) develop efficient methods to run large datasets, (2) use existing algorithms to bootstrap data, or build new simple algorithms as an initial bootstrap method, (3) consider using the convolutional neural net on signal datasets that already contain large numbers of samples (like North Atlantic right whale [NARW] sounds), and (4) leverage the research by participating in conferences and workshops.

### SYSTEM DEVELOPMENT

During Phase I the team devoted resources to the development of the DCL hardware and software infrastructure to serve as a testbench for processing large sound archives. The HPC system would be used throughout this project to support small to large data mining projects along with development of new strategies for tackling large, hard-to-process data. The backbone of this work was the development of an algorithm called the acoustic datamining accelerator algorithm (ADA), capable of dynamically assigning resources to distribute algorithms and sounds using parallel-processing technologies. In order to support large projects, a high performance-computing platform (HPC) was designed and constructed. The HPC tools were designed based on fielded systems [1-5] that offer a variety of desirable attributes, specifically dynamic resource allocation and scalability.

#### *HPC-ADA mini-cluster development*

A scalable hardware platform referred to as the HPC-ADA cluster, was developed as a powerful distributed server platform for executing big data applications for single, or multi-channel datasets. The HPC system developed for this work did not contain any unique or specialized hardware components. The main purpose for building the HPC-ADA systems was twofold; first keeping the hardware local gaurenteed integrity over the physical network providing a balanced configuration that was monitoried and controlled by researchers at Cornell, and secondly the sound archives were quite sizable; using other commercially available[1] systems would not be practical for moving sounds to remote servers for sound analysis and development. The HPC-ADA prototype system was constructed utilizing a DELL Cloud Server C6220, with remote access to other platforms. The system contains 64 physical cores of Intel Xeon E-2670 @ 2.6 GHz processor, 192 GB of local memory, 2 TB of local disk storage used for local cache, 64 TB of gigabit connected network attached storage (for sound archives and data products) and a 120 TB storage, hosting terrestrial and marine sound archives, connected at a campus level. The HPC-ADA server hosts the DeLMA[2] software application and other tools requuired for processing and manipulating large data projects.

[1] A possible extension of this work is bypass the HPC-ADA cluster and use Amazon EC2 cluster,both compatible with Mathworks MDCS.

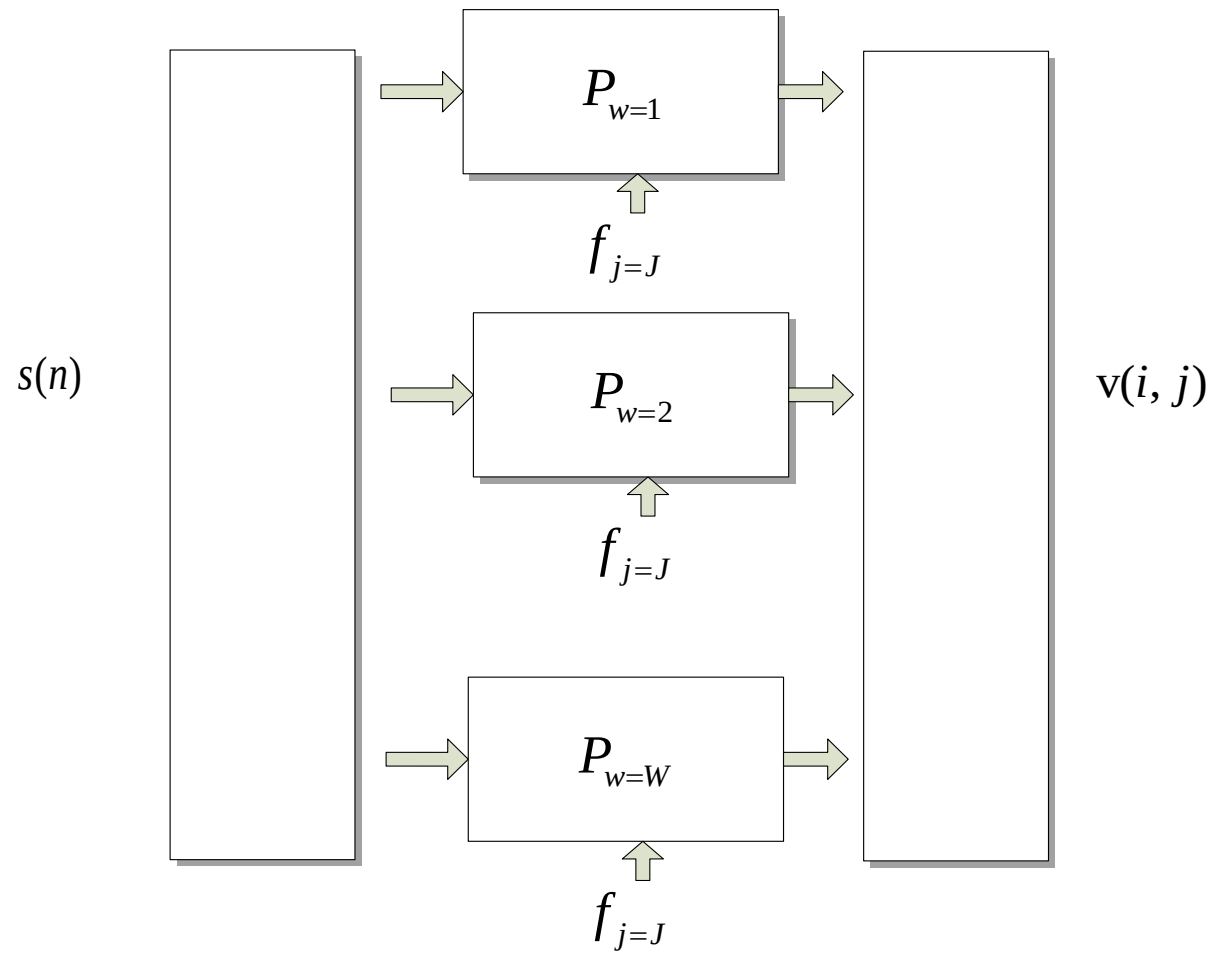

Figure 1. Parallel-distributed model for extracting data on archival sounds, scalable from small computers to large enterprise servers.

*ADA parallel-distributed algorithm*

The acoustic data-mining accelerator (ADA), was designed and implemented to provide a method for running detection-classification algorithms using parallel-distributed processing. The ADA technology was specifically designed for handling BIG sound archives and served as a runtime environment capable of distributing detection algorithms and sounds across parallel-distributed resources. Various projects utilized this technology since 2009, including both terrestrial and marine sources [31]. The grant provided funds to meaningfully enhance this technology with the capability to scale to large enterprise computer systems. Scaling from small to large was a major breakthrough for this technology. By using a single application running the ADA method, desktop computers or large enterprise systems can provide datamining capabilities through scalablity and concurrency, increasing throughput by several orders of magnitude. The ADA technology uses a complex mapping and gathering process as shown in Figure 1. The key to this process allowed DCL algorithms in their current serial format to be distributed across the resources. This point is shown in the figure as $f_j$, where each separate process (or computer core) is noted $P_w$; details for the work are published in [6, 7].

*User tools*

The ADA algorithm was reduced to practice using MATLAB base language accompanied by Mathworks Distributed Computing Server (MDCS), whereby the algorithm was integrated into a graphical user interface called DeLMA. The system incorporated the ADA algorithms and was designed to scale from a laptop (or desktop) application to a large, distributed systems. Design included a flexible HPC interface, capable of running a variety of algorithms concurrently across multiple datasets and sound formats. The output of the DeLMA engine was adapted to be viewed in RavenPro, a java application offered by Cornell[3] to the bioacoustic community (Figure 2). Significant work was done creating various tools for visualizing data using MATLAB routines. Tools include applications to view yearly distributional trends (diel response graphics), detection-classification performance metrics (ROC, DET and Precision-Recall curves), and feature space representations (such as cluster analysis). Much of the later work was done using prototype applications. The problem common to all these methods was in obtaining a standardized output format that would support large datasets and offer easy methods to sort and query information. The obvious solution was to attach a database friendly output to the DeLMA engine. Since this work was beyond the scope of the grant, significant work was put on hold until further funding is available. In the interim, however, basic applications were created for visualizing diel, detection performance and feature space results along with some basic performance studies for using a working database along with the companion application.

[2] Distributed sonic signal detection runtime using machine learning algorithms
[3] This refers to the Bioacoustics Research Program at the Cornell Lab of Ornithology.

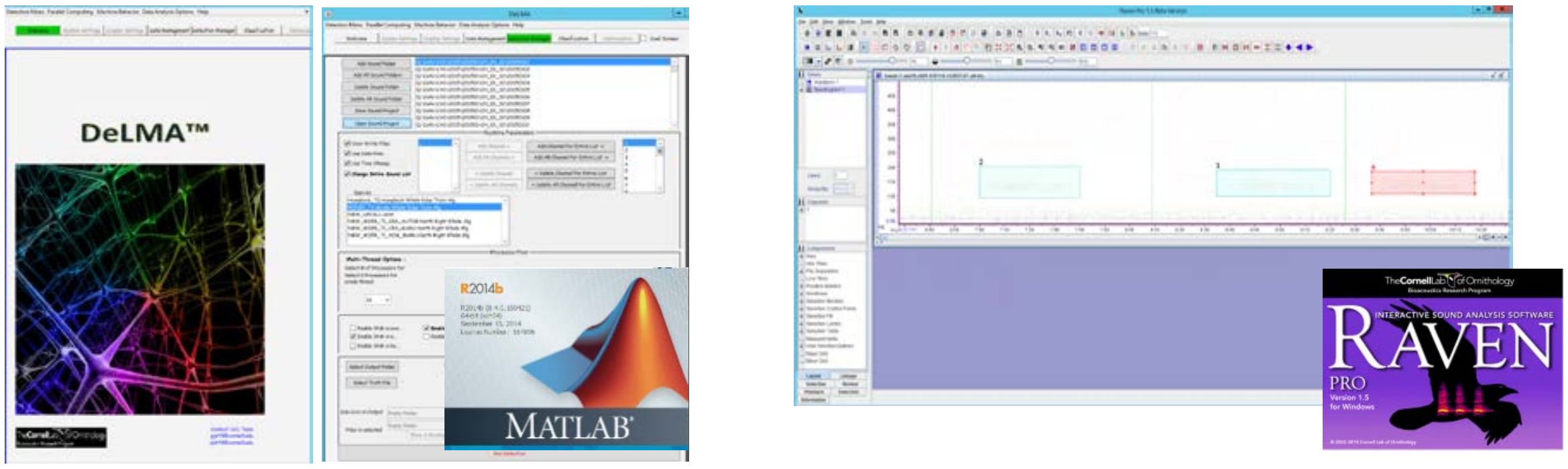

Figure 2. (left) MATLAB language used to build the ADA algorithm and incoporate into user interface, DeLMA. Data mining algrithms run using the DeLMA application on large sound archives, large datastes can be viewed by draging and droping DeLMA into RavenPro viewer (right).

*Data ouput format: Database benchmark tests*

Providing the user with easy methods to access the data for visualizing and analytical analysis resides in the output format. One concern for HPC processing resides in the amount of output data. Since the HPC can easily generate millions of records or events, management of the data becomes critical to useability. Our work tested four popular methods for measuring data management support. These methods were based on queries performed on the output data produced using the MATLAB DeLMA application. Formats include tab-delimited text file, matlab cell array, an XML database and SQL database. We used a simple schema based on popular detection event fields and constructed a dummy dataset of 100,000 events. The experiment measured the loading time and the query time for the data using each of the four methods. Performance from longest to shortest showed XML at 164.5 seconds, flat-file at 21.0 seconds, cell array at 10.9 seconds and the SQL at 2.4 seconds. Data extrapolated to 1 million events suggested a theoretical limit for XML to take 26 minutes, flat-file 3.6 minutes, cell array 1.8 minutes and the SQL about 24 seconds. In practice, theoretical limits do not account for a slow down due to system performance. For example, flat files with more than 1M events were not achievable in many actual cases; computer systems were not able to handle the file sizes. It is believed that a database format would offer a more scalable solution for larger projects.

## DATA-MINING ALGORITHM RESEARCH

*Algorithm integration and development*

Several advances were made for detection-classication algorithms. Since Cornell had several projects requiring detection classification work, existing algorithms were integrated with DeLMA-HPC and used as part of the analysis suite. Two popular methods were: (1) the multi-stage feature vector testing methods called isRAT [8], adopted to detect and recognize short duration, frequency-modulated (FM), tonal sounds from right whales and bowhead whales; and (2) a basic spectrogram correlation algorithm from xBAT [9] called the data-template, which uses an image correlation approach to recognize sample images (or templates). Some existing templetes were incorporated and used for detecting fin whale and bryde’s whale sounds, and mechanical noise. Both algorithm methods were integrated as part of the DeLMA-HPC suite, Table I.

Two new, advanced, multi-stage algorithms were investigated. The first method detects and charaterizes sequences of regularly repeated, similar acoustic events that we refer to as pulse-trains. The second major direction was the development of a method designed to achieve improved results for right whale up-call detection.

*Minke and seismic pulse-train algorithm*: Initial work developed an agorithm for detecting Atlantic Ocean minke whale songs [35] and was applied to data from a large-scale minke whale song project [10-14]. Since the technology was suited to detect pulse trains, the algorithm was modified and applied to other signals with similar characteristics. This included songs from Atlantic Ocean blue, fin, and humpback whales [35], as well as from impulses produced by seismic airgun array surveys [15]. To date, only the applications to minke and seismic airgun events have proven successful, while results for other signal types have shown positive results but need further invesitigation.

*Right whale algorithm:* The existing North Atlantic right whale (NARW) up-call detector lacked proper confidence scores and suffered from high false positive rates, making it difficult to use. With help from collaborators, the team sponsored an open competition for algorithm development in conjunction with an International Conference on Machine Learning (ICML) workshop. Two data competitions were leveraged through Kaggle.com and Marinexplore.com, collectively attracting 245 independent teams. Solutions varied, with several biologically inspired convolutional neural networks (CNN's) finishing near the top along with various hybrid algorithms. Scores were relatively high, with the top 10 entries having a mean score 98.0% +/- 0.25%. The competition also revealed, by comparing hand labeled results from two separate groups of analysts, that human truth labels, between each group, would disagree by as much as a 25% (+/-7.8%) within the same datasets [16]. Therefore any training done by a machine solution with 98% accuracy would be heavily biased by as much as 25% of the hand truth. Therefore, for purposes of implementation into DeLMA-HPC, BRP did not select the best scoring algorthm to implement into the HPC tools. Instead we considered three criterion; first, whether or not the algorithm had a simple code base; two, reltively easy to retrain, and three, had performance that exceeded 90% of the fielded solutions. The Cornell University solutions, referred to as the connected region analysis (CRA) and the histogram of oriented gradients (HOG) algorithms, finished in the top 20%, with 96.4% and 93.8% scores, respectively. Cornell subsequently selected both and implemented them in the DeLMA-HPC tools and applied them to a 12-month dataset. Results showed strong seasonal patterns for NARW calling behavior when running against large datasets [17]. The new algorithms have been adopted as an integral part of the right whale analysis protocol at Cornell and are currently being used on several projects.

*ASR: Architecture to support large data*

Early work indicated that signal segmenation is the single most important step in automatically recognizing whale sounds [12, 18, 19]. Other fields using recognition technologies adopted an approach that describes all the objects found in the scene using a single pass detection. There are several advantages to this approach, such as speed and context information. Feature extraction and classification can happen after successful segments (or regions) have been defined for the acoustic objects. In order to better optimize big data processing, a new approach called Acoustic Segmentation Recognition architecture (or ASR) was developed. [4, 5, 20-22]. A protoype of the ASR method was applied to the DeLMA software, and then two basic groups of signal types were tested. The first group consisted of short-tonal sounds called type-I, and the second group consisted of repeating short tonal sounds called type-II. The type-I basic routine was based on a connected region algorithm (CRA) and a series of rules to describe the signal [17, 23]. For example, *right whale* vocal calls were incorporated into the architecture, using connected region algorithm. The type-II routine was merely a repeating version of type-I.

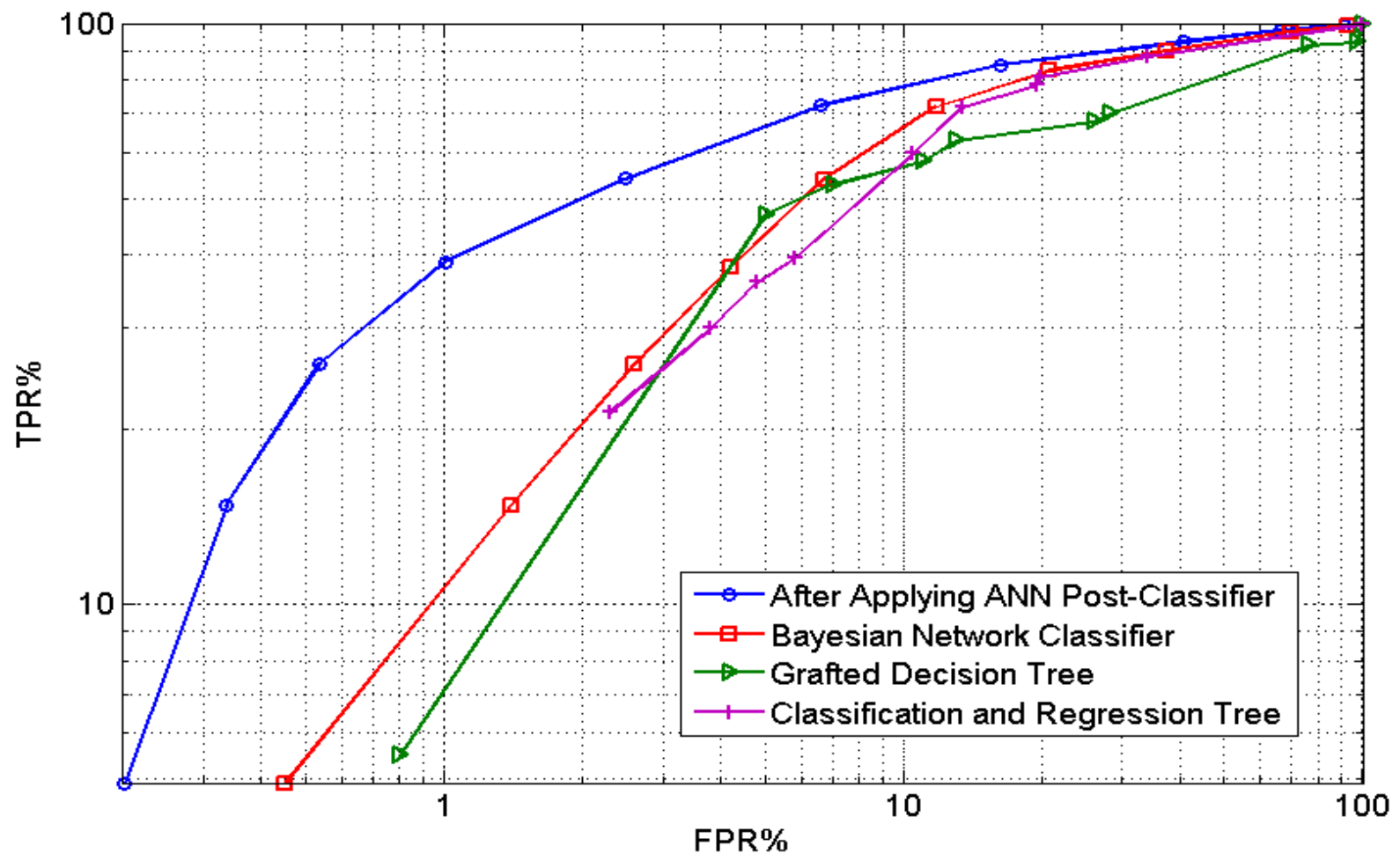

Figure 3 ROC of various common bioacoustic signal classifiers, and the effect of applying Human Knowledge Artificial Neural Network (HK-ANN) (Post-Classifier).

*Minke* pulse-train sounds were modeled by using CRA to describe the signal pulse with an added stage to measures the quality of the repeating pattern. Both type-I and type-II algorithms were developed and implemented to separate the stages according to the ASR format. Minke pulse train (type-II) detection-recognition performance was tested using the ASR structure; a total of 41,560 potential events (signal and noise) were generated. ASR provided a minimal number of events (< 3000) inspected and manually scored using "expert" human knowledge. A post classifier stage, trained and augmented the machine features and recognition accuracy. Figure 3 shows the Receiver Operating Curve (ROC) performance of the various common classifiers, the blue curve shows the performance of the system after applying the proposed post-classifier on the output. There is a demonstrable improvement in overall performance (especially at low False Positive Rates [FPR]), by using the new method. For example, at a FPR of 6% there is an approximately 20% improvement in True Positive Rate (TPR). Preliminary results show that combining human knowledge and machine features allowed for a post-classifier stage to require far less samples to learn from (< 3000) than the traditional multistage detection-classification methods. More importantly, having the "expert" in the loop offers an additional feature (or score) which can be considered "human judgement". By augmenting these metrics into the post-classification stage, a reasonably accurate result is achieved [6]. Further field testing for this work requires efficient data management (database tools) and remains for future work.

Table I. Popular data-mining algorithms that utilized HPC processing through this project.

| Algorithm Identifier | Species (Signal Type) | Algorithm description and reference. |
|---|---|---|
| 1 | Right Whale (up-call) | Custom multi-stage detection-recognition algorithm *isRAT*. [8, 24-26]. |
| 2<br>3 | Right Whale (up-call)<br>Elephant (pulse) | Detection-classification using histogram of oriented gradients [17, 19, 27]. |
| 4<br>5<br>6 | Seismic air gun (pulses)<br>Sperm, Minke, Fin (pulses)<br>Right Whale (up-call) | Multi-stage energy detection, using connected region analysis [11, 15, 23, 28]. |
| 7<br>8 | Brydes Whale (sweep)<br>Fin Whale (pulses) | Data-template and matched filtering concepts [9, 29]. |
| 9<br>10 | Minke Whale (pulse)<br>Fin Whale (pulse) | Multi-stage energy detection, connected region analysis and pulse-train, cross correlation [10, 11, 28, 30]. |

## COLLABORATIVE BIG DATA PROJECTS

### *Stellwagon Bank, NOPP sound archiving for online research*

This collaborative project focused on coordinating the mobilization of nearly four years of acoustic data recorded during an earlier NOPP project (N00014-07-1-1029). The focus was to mobilize a dataset consisting of several TeraBytes of data from 2006 to 2010. Data were recorded using Cornell MARU sensors deployed with different configurations and sample rates in the Stellwagen Bank National Marine Sanctuary (SBNMS) (see Table 2). This archive contains a rich collection of marine mammal sounds and served as a "sandbox" for our technology development. Sounds include a variety of anthropogenic noise sources, such as commercial vessels and fishing boats as well as a wide variety of whale (fin, humpback minke, right, and sei whales) and fish (haddock and cod) sounds. The archive offers a unique collection of sounds to promote development of HPC technologies and advanced data-mining algorithms. Publications generated in part from this work include [10, 11, 13, 31, 32].

### *JASCO and HARP data formats*

Cornell hosted several initiatives to expand the DeLMA and ADA components beyond the formats used for MARU-specific projects. In 2013, a noise analysis workshop was held at Cornell where participants brought datasets to process on the HPC machine. Tools were modified to accommodate the data organization (and format) supplied by JASCO and Cornell. Datasets consisted of large, continuous, multi-channel underwater recordings. Results for this work were captured in a white paper available through Cornell. In 2014 DeLMA-ADA software was adapted to support data from a single channel HARP sensor and from a large dataset consisting of 30 Tera Bytes of continous cable array recordings. In all cases, the DeLMA-HPC software was successfully modified to support the sound-librray formats and file organization. Formalized performance testing was published in [6]. Unpublished results for the HARP sensor showed a significant improvement in processing performance when using the HPC software in a multi-core, distributed configuration. Benchmarks for noise analysis computation from the HARP recording sensor spanning 28 day, 200 kHz, sound archive showed that a 64 core process was able to perform the computations and provide visual results in under 27 minutes; the same processing using a single core execution took over 359 minutes. For compatibility, other algorithms were tested on the HARP data and ran without error.

*Cornell data mining projects*

In 2015 Cornell inititated a large storage solution for hosting historical deployment, big data aspects of the work featured in several talks and a technical seminar [33-35]. The data store contains sounds from many Cornell marine and terrestrial projects wherein the HPC system was tested for access and processing performance across the campus networks. Projects include Massachusetts Bay (estimating 832k channel-hours), Gulf of Mexico (350k channel-hours), Baffin Bay (5.5k channel hours), Mass South (25k channel-hours), Gulf of Maine (26.3k channel-hours), Cape Cod Bay (21.6k channel-hours), SBNMS (60.4k channel-hours), Virginia Coast (23.5k channel-hours), and NAVFAC (East Coast) (10k channel-hours). The DeLMA HPC tools were used to process the sounds using various data-mining algorithms, see Table II. Ongoing work for Maryland and Virginia deployments is currently progressing.

Table II. Select projects that used HPC system and DeLMA runtime software.

| **Deployment** | **Channel Hours (Est.)** | **Job Runs** | **Algorithm [Signal Type ID[4]]** |
|---|---|---|---|
| Massachusetts Bay | 832k | 1 | Right Whale [1], Fin Whale [7] |
| Gulf of Mexico | 350k | 3 | Sperm Whale [5], Brydes Whale [9] |
| Baffin Bay | 5.5k | 5 | Seismic Air Gun [4], |
| Mass South | 25k | 3 | Minke Whale [6, 11], Right Whale [1, 2, 8] |
| Gulf of Maine | 26.3k | 2 | Minke Whale [11], Right Whale [1, 2, 8] |
| Cape Cod Bay | 21.6k | 6 | Right Whale [1, 2, 8], Minke Whale[11], Fin Whale [7] |
| SBNMS | 60.4k | 10 | Right Whale [1, 2, 8], Minke Whale [6, 11], Fin Whale [7] |
| Virginia | 23.5k | 2 | Right Whale [1, 2, 8], Minke Whale [6, 11], Fin Whale [7] |
| NAVFAC (32 kHz) | 10k | 2 | Sperm (PT), Right Whale [1, 2, 8], Minke Whale [11], Fin Whale [7, 12] |

## RESULTS

[4] Signal Type ID from Table I.

Systems and algorithms to process large data sets were designed and developed using COTS tools between 2011-2013. The development of the ADA algorithm software technology specialized to systematically distribute scalable computer resources was a critical development in this period. A working model of the ADA is contained in the user interface called DeLMA. The HPC technology developed has already successfully supported over 19 large projects at Cornell and executed over 3.6 million channel hours of analyzed sounds. Several commonly used algorithms were integrated into the technology, as were two newly developed algorithms designed for right whale and minke whale detections. The two new whale algorithms were assembled into a structure called ASR, allowing for post-processing data analytics (such as classification). Using a technique called the HK-ANN, results showed that users were able score and build a post-classifer using a relatively small number of acoustic objects resulting with as much as 20% improvement on TPR for a 6% FPR threshold. Final phases of the project showed positive results for processing large complex archives, for example, performance for 200kHz HARP datasets indicate dramatically improved runtimes of 13 times faster than conventional methods. These are promising results for supporting next-generation processing and analytics, particularly for challenging big datasets and high-bandwidth recorders.

The program accomplishments to date suggest several beneficial extensions. Since the ADA parallel-distributed algorithm is scalable, running on larger computers systems is relatively easy and would offer further benefits. Other algorithms within the accoustics community could also be integrated into the HPC framework such as whistle detection or general power law methods. Creating a relatively simple way to add algorithms to the tools and making these available through an agile software toolset would foster growth within the community and is key to the development and transferability of this work. Lastly, utilizing data management structures with advanced data mining methods (e.g., the HK-ANN method) would provide a more efficient approach to analyzing large stores of bioacoustic data than is currently offered by traditional methods applied to smaller datasets. Additional funding is critically important to progress this promising, ongoing research. Further funding will permit the development of inclusive technology to involve additional collaborators, thereby opening exploration of existing and newly acquired data. Increased and more comprehensive data analysis will undoubtedly expand the bioaccoustic community's understanding of animal ecology and biodiversity in the ocean.

## IMPACT/APPLICATIONS

The HPC-ADA machine and DeLMA software are models that have been successfully used with algorithms beyond detection-classification, including noise analysis and acoustic propagation modeling. Client-server architectures have also been explored through application of the MATLAB Distributed Computing Server (MDCS). Further applications and research for this work are ideal for leveraging systems requiring large, complex data-mining operations. Further investment for this work should be done at a larger scale within the bioacoustic community, offering access to a wider breadth and diversity of research scientists and analysis projects, with the additional goal of applied application to real-world situations.

## RELATED PROJECTS

Related and ongoing projects include continuing collaborations with the SBNMS and Marine Acoustics, Incorporated. Various internal efforts include projects supported by the Commonwealth of Massachusetts, and by BOEM off Virginia and Florida. Analyses requirements for publications from some earlier projects in the Gulf of Maine, Gulf of Mexico and off Florida continue to utilize processing capabilities developed from the HPC tools. A multi-group collaboration funded through the Synthesis of Arctic Research (SOAR) program included a work meeting held at Cornell and was heavily dependent on this project's system and tools developed [36]. Independent research through Marine Acoustics, Inc. continues work affiliated with the United States Navy. Integration with HARP sensor platform for supporting 200 kHz data was provided by John Hildebrand and Marie Roch, ongoing efforts for follow-on work is in proposal stages with ONR.

## PUBLICATIONS, PRESENTATIONS AND DATA COMPETITIONS

### PUBLICATIONS

PRESENTATIONS AND DATA COMPETITIONS

C.W.Clark, P.J.Dugan, Y. Le Cun, S. Van Parijs, D. Ponirakis and A. Rice, "Application of advanced analytics and high-performance-computing technologies for mapping occurrences of acoustically active marine mammals over ecologically meaningful scales," *key note talk*, ICML'13, Workshop on Machine Learning for Bioacoustics, Atlanta, Georgia, USA 2013.

P.J. Dugan and A. Rahaman, "Kaggle Competition, Cornell Univerity, The ICML 2013 Whale Challenge - Right Whale Redux, http://www.kaggle.com/c/the-icml-2013-whale-challenge-right-whale-redux, June 2013.

E. Spaulding, "Kaggle Competition, MarinExplore and Cornell University, Right Whale Challenge, https://www.kaggle.com/c/whale-detection-challenge, April 2013,".

M. Popescu, P. Dugan, J. Zollweg, A. Mikolajczyk and C.W Clark (2013), "Large-scale Detection and Classification (DC): Four Case Studies Using an Applied Distributed High Performance Computing (HPC) Platform, International Workshop on Detection Classification, Localization and Density Estimation (DCLDE), St. Andrews, Scotland.

P.J.Dugan, "High Performance Computing for Applied Detection Classification on Big-Acoustic Data", ERMITES Workshop, France 2013.

C.W. Clark, "Cornell Bioacoustics Scientists Develop a High-Performance Computing Platform for Analyzing Big Data", http://www.mathworks.com/company/user_stories/Cornell-Bioacoustics-Scientists-Develop-a-High-Performance-Computing-Platform-for-Analyzing-Big-Data.html, Mathworks Central, October 1, 2014.